\documentclass[graybox]{svmult}
\usepackage{ifthen}
\newboolean{arXiv}\setboolean{arXiv}{true}
\newboolean{proofing}\setboolean{proofing}{false}

\usepackage{type1cm}        
\usepackage{makeidx}         
\usepackage{graphicx}        
\usepackage{multicol}        
\usepackage[bottom]{footmisc}

\usepackage{newtxtext}       %
\usepackage{newtxmath}       

\usepackage{undertilde}

\makeindex             
\ifthenelse{\boolean{proofing}}{%
	\let\originalIndex=\index
\renewcommand{\index}[1]{\indexFormat{\originalIndex{#1}}}
\newcommand{\indexFormat}[1]{#1\makebox[0pt][c]{\smash{\raisebox{1.5ex}{$\color{red}\bullet$}}}}%
}{\relax} 

\makeatletter
\newcommand{\manuallabel}[2]{\def\@currentlabel{#2}\label{#1}}
\makeatother

\usepackage{enumitem}
\usepackage[breaklinks]{hyperref}
\usepackage{booktabs}
\usepackage{soul}
\usepackage[colorinlistoftodos,textsize=footnotesize]{todonotes}

\usepackage{amsmath,amssymb,mathbbol,mathrsfs,tikz}
\usetikzlibrary{matrix,arrows,patterns,cd}
   \usepackage{dsfont}

\DeclareMathOperator{\Var}{Var}
\newcommand{\CC}{\mathbb{C}}           

\newcommand{\Ocal}{\mathcal{O}}

\newcommand{\HH}{{\mathcal{H}}}

\newcommand{\If}{{\mathscr{I}}}

\newcommand{\II}{{\mathbb{1}}}

\newcommand{\Lb}{{\boldsymbol{L}}}
\newcommand{\Mb}{{\boldsymbol{M}}}
\newcommand{\Nb}{{\boldsymbol{N}}}

\newcommand{\Ac}{{\mathcal{A}}}
\newcommand{\Cc}{{\mathcal{C}}}
\newcommand{\Bc}{{\mathcal{B}}}

\newcommand{\Uc}{{\mathcal{U}}}

\newcommand{\id}{{\textrm{id}}}

\newcommand{\dvol}{{\textrm{dvol}}}

\newcommand{\CoinX}[1]{C_0^\infty(#1)}

\newcommand{\be}{\begin{equation}}
\newcommand{\ee}{\end{equation}}

\DeclareMathOperator{\Prob}{Prob}
\DeclareMathOperator{\supp}{supp}

\begin{document}
	\title*{A generally covariant measurement scheme for
		quantum field theory in curved spacetimes} 
	\titlerunning{Generally covariant measurement schemes for QFT in CST} 

\author{Christopher J Fewster}
\institute{CJ Fewster \at Department of Mathematics,
		University of York, Heslington, York YO10 5DD, United Kingdom \email{chris.fewster@york.ac.uk}}
 
	\maketitle 
	%
	%
	\abstract{%
	We propose and develop a measurement scheme for quantum field theory (QFT) in curved spacetimes, in which the QFT of interest, the ``system'', is dynamically coupled to another, the ``probe'', in a compact spacetime region. Measurements of observables in the probe system then serve as proxy measurements of observables in the system, under a correspondence which depends also on a preparation state of the probe theory. All our constructions are local and covariant, and the conditions may be stated abstractly in the framework of algebraic quantum field theory (AQFT). The induced system observables corresponding to probe observables may be localized in the causal hull 
	of the coupling region and are typically less sharp than the probe observable, but more sharp than the actual measurement on the coupled theory. A formula is given for the post-selected system state, conditioned on measurement outcomes, which is closely related to the notion of an instrument as introduced by Davies and Lewis. This formula has the important property that individual measurements form consistent composites, provided that their coupling regions can be causally ordered and a certain causal factorisation property holds for the dynamics; the composite is independent of the causal order chosen if more than one exists. The general framework is amenable to calculation, as is shown in a specific example. This contribution reports on joint work with R.~Verch,~{\tt arXiv:1810.06512.}} 
	\abstract*{%
	We propose and develop a measurement scheme for quantum field theory (QFT) in curved spacetimes, in which the QFT of interest, the ``system'', is dynamically coupled to another, the ``probe'', in a compact spacetime region. Measurements of observables in the probe system then serve as proxy measurements of observables in the system, under a correspondence which depends also on a preparation state of the probe theory. All our constructions are local and covariant, and the conditions may be stated abstractly in the framework of algebraic quantum field theory (AQFT). The induced system observables corresponding to probe observables may be localized in the causal hull 
	of the coupling region and are typically less sharp than the probe observable, but more sharp than the actual measurement on the coupled theory. A formula is given for the post-selected system state, conditioned on measurement outcomes, which is closely related to the notion of an instrument as introduced by Davies and Lewis. This formula has the important property that individual measurements form consistent composites, provided that their coupling regions can be causally ordered and a certain causal factorisation property holds for the dynamics; the composite is independent of the causal order chosen if more than one exists. The general framework is amenable to calculation, as is shown in a specific example. This contribution reports on joint work with R.~Verch,~{\tt arXiv:1810.06512.}} 
	%
	
	\section{Introduction}
	
	A typical first course on quantum mechanics will include some straightforward rules describing measurements. It might be said, for example, that an ideal measurement of
	a quantum mechanical observable returns one of the eigenvalues of the operator representing the observable, and the system is to be found in the corresponding eigenstate immediately afterwards, having changed instantaneously. Later, one learns that essentially every part of this rule is a considerable oversimplification, either technically or conceptually. Not least, the idea of an instantaneous transition is obviously problematic in relativistic theories. Quantum measurement theory\index{quantum measurement theory} (QMT) takes as its task the problem of refining rules of this type and putting them into an operational context. See~\cite{Busch_etal:quantum_measurement} for a recent comprehensive account. 
 	One strand of this work is the description of \emph{measurement schemes}\index{measurement scheme} that describe part of the process by which physical quantities may be measured. A system of interest is prepared, and then coupled to a probe, itself a quantum mechanical system, which is later measured. The probe result is then interpreted as a measurement of the system. The combination of the system, probe, coupling, and the probe observable measured, is called a measurement scheme for the system observable indirectly measured in this way. 
 	
 	This contribution describes joint work with Rainer Verch~\cite{FV:2018}, which
 	aims to adapt these ideas to describe measurements in quantum field theory in possibly curved spacetimes, using the methods of algebraic quantum field theory\index{algebraic quantum field theory} (AQFT)~\cite{Haag,Araki}. In so doing, we 
 	bridge a surprising gap in the literature: research on AQFT, despite its focus on local observables and local operations, has largely avoided the question of how such observables may be measured or operations performed; on the other hand, QMT has typically concentrated on quantum mechanical problems and avoided questions relating to spacetime localisation. Naturally, there are exceptions, and the works of Hellwig and Kraus~\cite{HellwigKraus:1969,HellwigKraus:1970,HellwigKraus_prd:1970} and more recent works~\cite{Ozawa:1984,OkamuraOzawa} may be mentioned as discussions in which questions of measurement in AQFT have been addressed. The reader will find much food for thought in the paper of Peres and Terno~\cite{PeresTerno:2004}. None of these, however, take quite the line that we describe here and, in particular, none of them consider the curved spacetime context. Again, this is a surprise, because of the prominence of the Unruh effect~\cite{Unruh:1976,CrispinoHiguchiMatsas:2008} in QFT in curved spacetimes (QFT in CST).
 	(See~\cite{BucVer_macroscopic:2015} for a recent discussion of some operational and interpretative aspects of the Unruh effect.) As mentioned, a full presentation of this work is given in~\cite{FV:2018}; our aim here is to present the outlines of the results obtained, while remaining reasonably self-contained. The main general questions considered are:
 	\begin{itemize}
 		\item what is the correspondence between probe observables and system observables?
 		\item in particular, what can be said concerning the spacetime localisation of the latter?
 		\item what is the appropriate description of state change following measurement, consistent with covariance?
 		\item how can one combine multiple observations in different spacetime regions, and the consequent changes of state, in a consistent way?
 	\end{itemize}
	We emphasise that the framework is on the one hand sufficiently broad to stand as a general description of measurement schemes in AQFT, while on the other it is sufficiently concrete to permit calculation in particular models; we will report the results of such calculations later on. To start, however, we begin by describing the nature of the system, probe and the coupling between them. 
 	
 	\section{System, probe, and coupling}
 	
 	In QMT the system and probe are usually quantum mechanical systems, with individual Hamiltonians $H_S$ and $H_P$ on respective Hilbert spaces $\HH_S$ and $\HH_P$. The
 	two can be treated as a single system with uncoupled Hamiltonian
 	\begin{equation}
 	H_U = H_S\otimes \II+ \II\otimes H_P
 	\end{equation}
 	on the combined Hilbert space $\HH=\HH_S\otimes\HH_P$. A coupling can be introduced and removed by modifying the Hamiltonian to 
 	\begin{equation}
 	H_C(t) = H_U + H_{\text{int}}(t)
 	\end{equation}
 	with $H_{\text{int}}(t)=0$ for sufficiently early and sufficiently late times $t$. Alternatively one might simply specify a unitary time evolution $U(t)$ on the combined Hilbert space, with $U(t)=\II$ for early and late $t$. The goal is then to understand
 	how measurements of observables on $\HH_P$ can be interpreted as measurements of observables on $\HH_S$. 
 	
 	In this section we will introduce analogous structures for two quantum field theories with a coupling confined to a compact region of spacetime. There are two problems. First, we wish to maintain covariance and therefore avoid introducing preferred time coordinates. This will be dealt with by adapting ideas from AQFT, particularly in the locally covariant description developed for curved spacetimes~\cite{BrFrVe03,FewVer:dynloc_theory}. Second, one might wonder about the physical status of coupling together quantum fields. After all, the interactions of nature are not ours to change -- what, then, is the relevance, of a formalism based on such modifications? The answer to this criticism is that the couplings represent a proxy for an experimental design that engineers interactions to occur in the apparatus and tries to screen out extraneous influences. For example, electromagnetism and QCD are coupled within the standard model. But if we probe the structure of a nucleus by directing a laser at it, we produce interactions in a particular spacetime region, and it becomes reasonable to neglect the interactions between electromagnetism and QCD elsewhere in spacetime. It is also true that by choosing interaction energies we may exploit the running of coupling constants to modify the strength of various interactions.
 	
 	We now explain the formalism in more detail, which requires a short discussion of AQFT in curved spacetimes. A recent introduction to AQFT in flat spacetimes can be found in~\cite{FewsterRejzner_AQFT:2019}\ifthenelse{\boolean{arXiv}}{\relax}{ (\hl{this volume})}; for an exposition of the locally covariant approach to QFT in CST, introduced in~\cite{BrFrVe03}, see~\cite{FewVerch_aqftincst:2015}.
 	
 	Recall that a time-oriented Lorentzian manifold spacetime $\Mb$ is globally hyperbolic\index{global hyperbolicity} if it contains no closed causal curves and each of its compact sets $K$ has a compact \emph{causal hull}\index{causal hull} $J^+(K)\cap J^-(K)$, where $J^{+/-}(S)$ represent the causal future/past\index{causal future/past} of a set $S$ (where needed, we will denote $J(S)=J^+(S)\cup J^-(S)$). Equivalently, $\Mb$ contains a Cauchy surface\index{Cauchy surface} -- a subset met exactly once by every inextendible timelike curve.
 	 A subset of $\Mb$ will be called \emph{causally convex}\index{causal convexity} if it is equal to its causal hull, 
 	which means that it contains every causal curve between any pair of its points. If $\Ocal$ is any open causally convex subset of a globally hyperbolic spacetime $\Mb$, 
 	then $\Ocal$, equipped with the induced metric and causal structure, is a globally hyperbolic spacetime in its own right, to be denoted $\Mb|_\Ocal$. 
 	
 	We shall be interested in QFTs that can be formulated on a class of globally hyperbolic spacetimes that is closed under the formation of subspacetimes in the manner just described. To each such spacetime $\Mb$, we assume that such the QFT is described on $\Mb$ by means of a unital $*$-algebra $\Ac(\Mb)$ and subalgebras $\Ac(\Mb;\Ocal)$ sharing the unit with $\Ac(\Mb)$ and labelled by open causally convex (but not necessarily precompact) subsets $\Ocal$ of $\Mb$. The assumptions needed here are
 \begin{enumerate}[label=\bf A\arabic{enumi},leftmargin=*,widest=4] 
 	\item\label{it:CSTiso} {\bf Isotony}\index{isotony}
 	Whenever $\Ocal_1\subset \Ocal_2\subset\Mb$, the corresponding local algebras are nested, 
 	\begin{equation}
 	\Ac(\Mb;\Ocal_1)\subset \Ac(\Mb;\Ocal_2).
 	\end{equation} 
 	\item\label{it:CSTEins} {\bf Einstein causality}\index{Einstein causality} If $\Ocal_1$ and $\Ocal_2$ are causally disjoint subsets of $\Mb$ then $\Ac(\Mb;\Ocal_i)$ commute element-wise,
 	\begin{equation}
 	[\Ac(\Mb;\Ocal_1),\Ac(\Mb;\Ocal_2)] = 0.
 	\end{equation}
 	\item\label{it:CSTcomp} {\bf Compatibility}\index{compatibility} If $N$ is an open causally convex subset of $\Mb$, then there is an injective unit-preserving $*$-homomorphism (to be described henceforth as a \emph{monomorphism}\index{monomorphism}) $\alpha_{\Mb;\Nb}:\Ac(\Nb)\to\Ac(\Mb)$ whose image is precisely $\Ac(\Mb;N)$, where $\Nb=\Mb|_N$. We will refer to $\alpha_{\Mb;\Nb}$ as a \emph{compatibility map}\index{compatibility map}. Whenever $M_3\subset M_2\subset M_1$ the compatibility maps obey the relation
 	\begin{equation}\label{eq:functorial}
 	\alpha_{\Mb_1;\Mb_2}\circ \alpha_{\Mb_2;\Mb_3} = \alpha_{\Mb_1;\Mb_3}.
 	\end{equation} 
 	\item\label{it:CSTdyn} {\bf Timeslice property (existence of dynamics)}\index{timeslice property}
 	If $N$ is an open causally convex subset of $\Mb$ and contains at least one Cauchy surface of $\Mb$, then $\alpha_{\Mb;\Nb}$ is an isomorphism; equivalently, $\Ac(\Mb;N)=\Ac(\Mb)$. 
 	In conjunction with compatibility, we may deduce that whenever
 	$\Ocal_1\subset \Ocal_2\subset\Mb$ and $\Ocal_1$ contains a Cauchy surface of $\Ocal_2$, then 
 	\begin{equation}
 	\Ac(\Mb;\Ocal_1) = \Ac(\Mb;\Ocal_2).
 	\end{equation} 
 	\item\label{it:CSTHaagprop} {\bf Haag property}\index{Haag property} Let $K$ be a compact subset of $\Mb$. Suppose that $A\in\Ac(\Mb)$ commutes with every element of $\Ac(\Mb;N)$ for every region $N$ contained in the causal complement\index{causal complement} $K^\perp=\Mb\setminus J(K)$ of $K$. Then   $A\in\Ac(\Mb;L)$ whenever $L$ is a \emph{connected} open causally convex subset containing $K$. This is a weakened form of Haag duality\index{Haag duality}~\cite{Haag}. 
 \end{enumerate}
 	Comparing with assumptions standard in flat spacetime (see, e.g.,~\cite{FewsterRejzner_AQFT:2019}) one notes that Poincar\'e covariance has been replaced by the compatibility condition. What we have described is a cut-down version of
 	locally covariant QFT~\cite{BrFrVe03,FewVerch_aqftincst:2015}, the full version of which would describe the theory as a functor from the category of globally hyperbolic spacetimes to a category of unital $*$-algebras. Einstein causality may be built in by suitable monoidal~\cite{BrFrImRe:2014} or operadic~\cite{BeniniSchenkelWoike:2019} refinements. However we will not need this level of generality here. On a point of terminology, if $A\in\Ac(\Mb;\Ocal)$ we will say that $A$ is \emph{localisable}\index{localisable algebra element} in $\Ocal$, or that $\Ocal$ is a \emph{localisation region}\index{localisation region} for $A$; due to~\ref{it:CSTiso} and~\ref{it:CSTdyn} there will be many possible localisation regions for a given $A$. Finally, recall that a state\index{state} of a theory is a positive normalised linear map $\omega:\Ac(\Mb)\to\CC$, and the expectation value\index{expectation value} of observable $A\in\Ac(\Mb)$ in state $\omega$ is $\omega(A)$. 
 	
 	Let two theories $\Ac$ and $\Bc$ be given, each of which obeys the above axioms, with 
 	compatibility maps $\alpha_{\Mb;\Nb}$ and $\beta_{\Mb;\Nb}$ respectively. We will regard $\Ac$ as describing the system and $\Bc$ the probe. 
 	Then we obtain a further theory $\Uc$, by defining 
 	\begin{equation}
 	\Uc(\Mb)=\Ac(\Mb)\otimes\Bc(\Mb), 
 	\end{equation}
 	for each spacetime $\Mb$, and using  $\upsilon_{\Mb;\Nb}=\alpha_{\Mb;\Nb}\otimes\beta_{\Mb;\Nb}:\Uc(\Nb)\to\Uc(\Mb)$ as the compatibility map when $N$ is an open causally convex subset of $\Mb$. This is the analogue of the tensor product of the system and probe in quantum mechanics; as ever in AQFT, the primary focus is on algebras of observables rather than on Hilbert spaces. 
 	
 	We wish to describe a coupling between the system and probe in an abstract way, without needing a specific Lagrangian description of the theories involved, and also without introducing privileged coordinates. Let $K$ be a compact subset of spacetime $\Mb$. Then the regions $M^\pm = \Mb\setminus J^\mp(K)$ are causally convex and open and constitute covariantly defined `in' ($-$) and `out' ($+$) regions, so that the `out' region has no intersection with the causal past of $K$ and the `in' region has no intersection with the causal future of $K$. 
 	\begin{definition}
 		A theory $\Cc$, with compatibility maps $\gamma_{\Mb;\Nb}$, is a \emph{coupling of $\Ac$ and $\Bc$ within $K$}\index{coupled theory, abstract definition of} if, there is a family of isomorphisms 
 		\begin{equation}
 		\chi_L:\Uc(\Lb)\to \Cc(\Lb)
 		\end{equation}
 		labelled by the open causally convex subsets $L$ of $M\setminus(J^+(K)\cap J^-(K))$,   with the following compatibility property: for any two such subsets $L, L'$ with $L'\subset L$, one has a commuting diagram
  		\begin{equation}\label{eq:naturality}
 		\begin{tikzcd}[column sep=large]
 		\Uc(\Lb') \arrow[r, "\upsilon_{\Lb;L'}"]\arrow[d,"\chi_{\Lb'}"] &
 		\Uc(\Lb) \arrow[d,"\chi_\Lb"]  \\
 		\Cc(\Lb') \arrow[r, "\gamma_{\Lb;L'}"] & \Cc(\Lb)
 		\end{tikzcd}. 
 		\end{equation}
 		\end{definition}
  		In short, this definition asserts that the theories $\Uc$ and $\Cc$ coincide outside the causal hull of $K$, capturing the idea that the coupling is only switched on in this region. There is a close link to the discussion of equivalence between theories in local covariant QFT~\cite{BrFrVe03,FewVer:dynloc_theory,FewVerch_aqftincst:2015}.  
  		
  		Many properties of the coupling can be described in terms of a scattering map defined as follows. For brevity, we denote the compatibility maps $\alpha_{\Mb;M^\pm}$ associated with the regions $M^\pm$ by $\alpha^\pm$, and the isomorphisms $\chi_{\Mb^\pm}$ by $\chi^\pm$ (by construction, $M^\pm$ are indeed open causally convex subsets not intersecting the causal hull of $K$). By the timeslice property, the monomorphisms
  		$\alpha^\pm$, $\beta^\pm$, $\upsilon^\pm$, $\gamma^\pm$, are isomorphisms, so the same applies to the compositions
  		\begin{equation}
  		\kappa^\pm = \gamma^\pm\circ \chi^\pm :\Uc(\Mb^\pm)\to \Cc(\Mb),
  		\end{equation}  
  		and also to the \emph{retarded ($+$) and advanced ($-$) response maps}\index{retarded/advanced response} 
  		\begin{equation}
  		\tau^\pm = \kappa^\pm\circ(\upsilon^\pm)^{-1}:\Uc(\Mb)\to\Cc(\Mb),
  		\end{equation}
  		and finally to the \emph{scattering morphism}\index{scattering morphism}
  		\begin{equation}\label{eq:Theta}
  		\Theta= (\tau^{-}){}^{-1}\circ \tau^+:\Uc(\Mb)\to\Uc(\Mb),
  		\end{equation} 
  		which maps algebra elements from late to early times. The scattering morphism has important properties~\cite[Prop.~1]{FV:2018}: in particular, it is unchanged if the notional coupling region $K$ is expanded, but keeping $\Cc$ the same; it also acts trivially on any subalgebra $\Ac(\Mb;L)$ labelled by a subset $L\subset K^\perp$. 
  		
  		\section{Induced system observables}
  		
  		The measurement scheme may be described loosely as follows. At early times, the system and probe are separately prepared in states $\omega$ and $\sigma$ respectively. They are then coupled; finally, at late times, a probe observable $B\in\Bc(\Mb)$ is measured. 
  		This description is somewhat imprecise because the actual world of the experiment is described by the coupled theory $\Cc$, rather than the separate theories $\Ac$ and $\Bc$, or their uncoupled combination $\Uc$. A little more precisely, what is meant is that, at early times, one prepares a state of $\Cc(\Mb)$ that has no correlations between system and probe degrees of freedom, and at late times an observation is made that only tests degrees of freedom in the probe. The problem of translating statements about the coupled theory into the language of the uncoupled theory is solved by the response maps. 
  		
  		Specifically, the state of $\Cc(\Mb)$ corresponding at early times to the product state
  		$\omega\otimes\sigma$ of $\Uc(\Mb)$ is given by
  		\begin{equation}
  		\utilde{\omega}_\sigma =  (\tau^-){}^{-1*}(\omega\otimes\sigma)
  		\end{equation}
  		where the star denotes the adjoint map, i.e., $\utilde{\omega}_\sigma(C)=(\omega\otimes\sigma)(\tau^- C)$. Likewise, the observable
  		$B\in\Bc(\Mb)$, may be identified with $\II\otimes B\in \Uc(\Mb)$ and hence corresponds to the observable  
  		\begin{equation}\label{eq:Btilde}
  		\widetilde{B}=  \tau^+(\II\otimes B)
  		\end{equation}
  		of $\Cc(\Mb)$, with the identification being made at late times. The expected measurement outcome from the experiment is the expectation value for the observable $\widetilde{B}$ in the state $\utilde{\omega}_\sigma$, which may be written
  		\begin{equation}\label{eq:probe_exp}
  		\utilde{\omega}_\sigma(\widetilde{B}) = (\omega\otimes\sigma)(\Theta(\II\otimes B)).
  		\end{equation}  
  		Notice that we use the advanced response to identify the states of the uncoupled and coupled systems at early times, and the retarded response to identify observables at late times. This reflects the fundamental time-asymmetry of the measuring process,\index{time asymmetry in measurement} which may be sloganized as \emph{prepare early and measure late}. 
  		
  		The goal is to interpret $\utilde{\omega}_\sigma(\widetilde{B})$ as the expectation value of a system observable in state $\omega$. To this end, we introduce the map $\eta_\sigma:\Ac(\Mb)\otimes\Bc(\Mb)\to \Ac(\Mb)$ defined by
  		\begin{equation}
  		\eta_\sigma(A\otimes B)=\sigma(B)A,
  		\end{equation}
  		extended by linearity, which is a completely positive map closely related to a conditional expectation.\index{conditional expectation} Then the map   $\varepsilon_\sigma:\Bc(\Mb)\to\Ac(\Mb)$ defined by 
  		\begin{equation}\label{eq:inducedobs}
  		\varepsilon_\sigma(B)=(\eta_\sigma\circ\Theta)(\II\otimes B) 
  		\end{equation} 
  		clearly satisfies  	
  		\begin{equation}
  		\omega(\varepsilon_\sigma(B)) = \omega((\eta_\sigma\circ\Theta)(\II\otimes B))
  		= (\omega\otimes\sigma)(\Theta(\II\otimes B)) = \utilde{\omega}_\sigma(\widetilde{B}) ,
  		\end{equation}
  		which provides an interpretation of the measurement in terms of the \emph{induced system observable}\index{induced system observable} $\varepsilon_\sigma(B)$. In other words, we have a measurement scheme for $\varepsilon_\sigma(B)$. The following is proved as~\cite[Thm~2]{FV:2018} 
  		\begin{theorem}\label{thm:induced}
  			For each probe preparation state $\sigma$, $A=\varepsilon_\sigma(B)$ is the unique $\omega$-independent solution to $\omega(A)=\utilde{\omega}_\sigma(\widetilde{B})$, provided that $\Ac(\Mb)$ is separated by its states. In general, the map $\varepsilon_\sigma$ is a completely positive linear map with the properties
  			\begin{equation}\label{eq:cpprops}
  			\varepsilon_\sigma(\II)=\II,\qquad \varepsilon_\sigma(B^*)=\varepsilon_\sigma(B)^*, \qquad
  			\varepsilon_\sigma(B)^*\varepsilon_\sigma(B)\le \varepsilon_\sigma(B^*B) \,.
  			\end{equation} 	
  			For fixed $B$, the map $\sigma\mapsto \varepsilon_\sigma(B)$ is weak-$*$ continuous.
  		\end{theorem} 
  		An immediate consequence is that each self-adjoint $B=B^*\in\Bc(\Mb)$ is mapped to a
  		self-adjoint element $\varepsilon_\sigma(B)$ of $\Ac(\Mb)$. It is important to note that 
  	    $\varepsilon_\sigma$ is generally neither an injection, nor an algebra homomorphism.   
  		Indeed, the inequality in~\eqref{eq:cpprops} implies
  		\begin{equation}
  		\Var (\widetilde{B};\utilde{\omega}_\sigma)  \ge  \Var(\varepsilon_\sigma(B);\omega),
  		\end{equation}
  		that is, the variance of measurement results in the actual experiment of measuring
  		$\widetilde{B}$ in state $\utilde{\omega}_\sigma$ is at least the variance of the 
  		hypothetical equivalent measurement of $\varepsilon_\sigma(B)$ in state $\omega$. The failure of $\varepsilon_\sigma$ to be a homomorphism is intimately connected to the existence of experimental noise, i.e., fluctuations of the probe. As another illustration, if $B$ is a projection, then $\varepsilon_\sigma(B)$ is in general not a projection, but rather an \emph{effect}\index{effect}; that is, an algebra element $A$ so that both $A$ and $\II-A$ are positive, i.e., $0\le A\le \II$. In general, 
  		an effect $A$ models an unsharp zero-one measurement, with
  		\begin{equation}
  		\Prob(A=1\mid \omega) = \omega(A), \qquad \Prob(A=0\mid \omega) = \omega(\II- A)
  		\end{equation} 
  		in state $\omega$. Thus, 
  		sharp (projective) measurements of the probe are to be interpreted as unsharp measurements\index{unsharp measurement} of the system. Combining our two observations, it may be said that
  		$\varepsilon_\sigma(B)$ is typically less sharp than $B$, but more sharp than the actual experimental observation. 
  		
  		An important question concerns the localisation of the induced system observables, which turns on the locality properties of the scattering morphism. Let $L$ be an open causally convex subset of $K^\perp$, so $\Theta$ acts trivially on $\Uc(\Mb;L)$. This has the following consequences. First, if $B\in\Bc(\Mb;L)$, we have $\II\otimes B\in\Uc(\Mb;L)$ and hence the corresponding induced system observable is 
  		\begin{equation}
  		\varepsilon_\sigma(B)=\eta_\sigma(\Theta (\II\otimes B)) = \eta_\sigma(\II\otimes B) = 
  		\sigma(B)\II.
  		\end{equation} 
  		As one would hope, no information concerning the system can be obtained by measuring a probe observable at spacelike separation from the coupling region; our framework does not allow for nonlocal signalling. 
  		 
  		Second, suppose that $A\in \Ac(\Mb;L)$ be any system observable localised in $L$. Then for an arbitrary probe observable $B\in\Bc(\Mb)$ one has 
  		\begin{align}
  		[\varepsilon_\sigma(B) ,A] &= [\eta_\sigma(\Theta (\II\otimes B)),A] 
  		= \eta_\sigma([\Theta (\II\otimes B),A\otimes \II)]) \\
  		&= \eta_\sigma(\Theta [\II\otimes B,A\otimes \II]) 
  		=0,\nonumber
  		\end{align}
  		as $\Theta$ leaves $A\otimes \II$ invariant. By the Haag property, we infer that $\varepsilon_\sigma(B)$ may be localised in any connected open causally convex set containing the coupling region $K$ (and hence containing its causal hull). We have proved:
  		\begin{theorem}[{\cite[Thm 3]{FV:2018}}]\label{thm:localisation}
  			For each probe observable $B\in\Bc(\Mb)$, the induced system observable $\varepsilon_\sigma(B)$ may be localised in any connected open causally convex set containing $K$. If $B$ may be localised in $K^\perp$ then $\varepsilon_\sigma(B)=\sigma(B)\II$.
  		\end{theorem}  
  		
  		The failure of $\varepsilon_\sigma$ to be a homomorphism is natural from another perspective. Suppose two incompatible (noncommuting) system observables $A_i$ ($i=1,2$) are localisable in the causal hull of $K$ and are in fact induced by probe observables $B_i$,  $A_i=\varepsilon_\sigma(B_i)$. If $\varepsilon_\sigma$ were a homomorphism, we would conclude that the $B_i$ are incompatible, which (by Einstein causality) would prohibit the possibility of localising them in spacelike separated regions. Turning this around, probe measurements in spacelike separated regions that correspond to the   $A_1$ and $A_2$ provide an unsharp joint measurement\index{unsharp joint measurements} of incompatible system observables.

  	\section{Instruments and change of state}
  	
  	Suppose an effect $B$ of the probe is measured, resulting in the value $1$ (we also say that the effect has been observed). How should the system state $\omega$ be updated in consequence? One way of addressing this question is to require that the updated state $\omega'$ on $\Ac(\Mb)$ should have the property that
  	\begin{equation}\label{eq:postselection}
  	\Prob(A; \omega') = \Prob_\sigma(A\mid B; \omega)= \frac{\Prob_\sigma(A\& B;\omega)}{\Prob_\sigma(B;\omega)}
  	\end{equation}
  	for all system effects $A$ measured at late times. Here, the middle quantity is the classical conditional probability\index{conditional probability}, in the original state, that the effect $A$ is observed, given that $B$ is also observed, with the subscript indicating the likely dependence on the probe preparation state $\sigma$. On the right-hand side, $A\& B$ is the joint effect\index{effect!joint} corresponding to the operator $A\otimes B\in\Uc(\Mb)$. By the reasoning used in the previous section, we have
  	\begin{equation}
  	\Prob_\sigma(A\& B;\omega) = \omega(\eta_\sigma\Theta (A\otimes B)), \qquad 
  	\Prob_\sigma(B;\omega) = \omega(\eta_\sigma\Theta (\II\otimes B))
  	\end{equation}
  	and so our requirement on $\omega'$ is
  	\begin{equation}\label{eq:IfIf}
  	\omega'(A) = \frac{(\If_\sigma(B)(\omega))(A)}{(\If_\sigma(B)(\omega))(\II)},
  	\end{equation}
  	where 
  	\begin{equation}\label{eq:instrument}
  	(\If_\sigma(B)(\omega))(A):= (\omega\otimes\sigma)(\Theta(A\otimes B)) = (\Theta^*(\omega\otimes\sigma))(A\otimes B)
  	\end{equation}
  	and the forest of parentheses indicates that $\If_\sigma(B)$ is to be regarded as a map
  	on the dual space of $\Ac(\Mb)$. 
  	Assuming that the algebra separates the states, equation~\eqref{eq:IfIf} implies that  
  	\begin{equation}
  	\omega':=\frac{\If_\sigma(B)(\omega)}{\If_\sigma(B)(\omega)(\II)}
  	\end{equation} 
  	is the \emph{post-selected system state}\index{post-selected state} after selective measurement of the probe.\footnote{Actually, one must check that $\omega'$ is indeed a state; see~\cite{FV:2018}.} The term `post-selection' is used in various ways in the literature; to be clear, the meaning intended here is given by~\eqref{eq:postselection}, 
  	i.e., the post-selected state is the one in which the probability of observing a system effect $A$ is equal to the conditional probability, in the original state, of observing $A$  given that the probe effect is observed. 
  	
  	The \emph{pre-instrument}\index{instrument!pre-instrument} $\If_\sigma(B)$ defined by~\eqref{eq:instrument} is a positive map on the dual space $\Ac(\Mb)^*$; that is, it maps positive linear functionals to positive linear functionals. 
  	Davies and Lewis~\cite{DaviesLewis:1970} introduced the term \emph{instrument}\index{instrument} to describe a measure on the $\sigma$-algebra of measurement outcomes valued in positive maps $\Ac(\Mb)^*$;
  	our pre-instruments are related to instruments as follows: given any measure ${\sf E}$ valued in the effects of $\Ac(\Mb)$, then $X\mapsto \If_\sigma({\sf E}(X))$ is a (completely positive) instrument in the Davies--Lewis sense. 
  		
  	The discussion above includes \emph{non-selective probe measurement}\index{non-selective measurement} as the special case $B=\II$, because the absence of any filtering on the measurement outcome is equivalent to employing a probe effect that is observed with probability $1$ in all states. Therefore the updated state resulting from the non-selective measurement is
  		\begin{equation}
  		\omega'_{\text{ns}}(A) =  \If_\sigma(\II)(\omega)(A) = (\Theta^*(\omega\otimes\sigma))(A\otimes \II),
  		\end{equation}
  	which is the partial trace of the state $\Theta^*(\omega\otimes\sigma)$ over the probe.
  	By the locality properties of $\Theta$, it follows immediately that $\omega'_{\text{ns}}(A) = \omega(A)$ if $A\in\Ac(\Mb;K^\perp)$. This shows that a non-selective measurement cannot influence the results of measurements in causally disjoint regions. The general situation for selective probe measurements is the following~\cite[Thm 4]{FV:2018}.
  	\begin{theorem}\label{thm:cor}
  		Consider a measurement of a probe effect $B$ in which the effect is observed. For each $A\in\Ac(\Mb;K^\perp)$, the expectation value of $A$ is unchanged in the post-selected state $\omega'$ if and only if $A$ is uncorrelated with $\varepsilon_\sigma(B)$ in the original system state $\omega$, i.e., $\omega(A\varepsilon_\sigma(B))=\omega(A)\omega(\varepsilon_\sigma(B))$.
  	\end{theorem}
  	This includes our non-selective result because $\varepsilon_\sigma(\II)=\II$ is uncorrelated with every system observable in every state. In general, updating the state by post-selection changes expectation values at spacelike separation from the measurement region (and, one expects, also in its past and future). Theorem~\ref{thm:cor} shows that the explanation is simply to do with correlation and inference. To give a non-quantum, non-relativistic example: if two envelopes, one red and the other blue, are sent through the post to two addresses, then the recipient of the red envelope may infer that the envelope at the other address is blue. 
  	
  	It may be clear that our account leans towards, though is not predicated upon, the view that the state is not a physical entity,\footnote{`[A] wavefunction is not a physical object. It is only a tool for computing the probabilities of objective macroscopic events' -- Peres and Terno in~\cite{PeresTerno:2004}.} and away from attempts to extend ideas like the `instantaneous collapse of the wavefunction' to QFT (cf.~\cite{HellwigKraus_prd:1970}). Particularly if one adopts a more subjective view on the nature of the state, there is a significant question of consistency to be addressed. Experiments conducted by multiple students in mutually spacelike separated laboratories might comprise a single experiment controlled by their supervisor. How can the updated states obtained by the students be reconciled, covariantly, with the overall update made by the supervisor? 
  	
  	Suppose that two probes are coupled to the system in compact interaction regions $K_1$ and $K_2$, so that no point of $K_2$ lies to the past of any point in $K_1$. At least some observers regard $K_2$ as occurring later than $K_1$, though it is not excluded that some might regard $K_2$ as earlier than $K_1$ (which happens if $K_1$ and $K_2$ are causally disjoint). To each probe system $\Bc_i$ there is a coupled system $\Cc_i$ and a scattering morphism $\Theta_i$ on the uncoupled algebra $\Uc_i(\Mb)=\Ac(\Mb)\otimes\Bc_i(\Mb)$. On the three-fold tensor product $\Ac(\Mb)\otimes\Bc_1(\Mb)\otimes\Bc_2(\Mb)$, these scattering maps
  	may be represented by 
  	\begin{equation}
	 \hat{\Theta}_1=\Theta_1\otimes_3 \id_{\Bc_2(\Mb)}, \qquad \hat{\Theta}_2=\Theta_2\otimes_2 \id_{\Bc_1(\Mb)},
  	\end{equation}
  	where the subscript on the tensor product indicates the slot into which the identity is inserted in each case. Alternatively, the two probes may be described as a single probe with algebra $\Bc_1(\Mb)\otimes\Bc_2(\Mb)$, coupling region $K_1\cup K_2$ and a combined
  	scattering morphism $\hat{\Theta}$ on $\Ac(\Mb)\otimes\Bc_1(\Mb)\otimes\Bc_2(\Mb)$.
  	A natural assumption is that the \emph{causal factorisation formula}\index{causal factorisation formula}
  	\begin{equation}\label{eq:Bogoliubov}
  	\hat{\Theta} = \hat{\Theta}_1\circ\hat{\Theta}_2
  	\end{equation}
  	holds -- this can be checked in examples but is expected on the basis e.g., of 
  	Bogoliubov's factorisation formula in perturbative QFT~\cite{Rejzner_book,DuetschFredenhagen:2000,BogoliubovShirkov}. Note that the map $\hat{\Theta}_2$ appears to the right because our scattering morphisms map from the future to the past. One of the main results of~\cite{FV:2018} is the following consistency result. 
  	\begin{theorem}[{\cite[Thm 5]{FV:2018}}]\label{thm:causalcomp}
  		Consider two probes as described above, with $K_2\cap J^-(K_1)=\emptyset$. For all probe preparations $\sigma_i$ of $\Bc_i(\Mb)$ and all probe observables $B_i\in\Bc_i(\Mb)$, the following identity for the pre-instruments holds:
  		\begin{equation}\label{eq:composite_instrument1}
  		\If_{\sigma_2}(B_2)\circ \If_{\sigma_1}(B_1) = \If_{\sigma_1\otimes \sigma_2}(B_1\otimes B_2).
  		\end{equation}
  		If, in fact, $K_1$ and $K_2$ are causally disjoint, we have
  		\begin{equation}\label{eq:composite_instrument2}
  		\If_{\sigma_2}(B_2)\circ \If_{\sigma_1}(B_1) = \If_{\sigma_1\otimes \sigma_2}(B_1\otimes B_2)  = \If_{\sigma_1}(B_1)\circ \If_{\sigma_2}(B_2).
  		\end{equation}
  	\end{theorem}
  	The reader is referred to~\cite{FV:2018} for the proof, which is a direct calculation. 
  	Suppose that $\omega_1(B_1)\neq 0$, so there is a nonzero probability of observing $B_1$, and that $\omega_1'(B_2)\neq 0$, where $\omega_1'$ is the updated state conditioned on the observation of $B_1$. Then another direct calculation shows that the post-selected state $\omega''_{12}$ conditioned on the observation of $B_2$ in state $\omega'_1$ coincides with the post-selected state $\omega'_{12}$ conditioned on the combined effect $B_1\otimes B_2$ being observed in state $\omega$. In other words, the updating may be made in stages, and in any order if the $B_i$ have causally disjoint localisation. No additional assumptions are needed beyond those we have mentioned; no geometric boundaries across which state reduction occurs are needed. The cornerstone is locality of the interactions in the coupled theories, expressed through~\eqref{eq:Bogoliubov}.  	
  		
  		\section{A specific model}\label{sec:model}
  		
  		The general formalism described above is amenable to practical calculation~\cite{FV:2018}. 
  		For instance, consider a simple situation in which both the system and the probe
  		are quantized real scalar fields of possibly different mass, with classical action
  		\begin{equation}
  		S_0 = \frac{1}{2}\int_M \dvol \left((\nabla_a \Phi)(\nabla^a\Phi) - m_\Phi^2 \Phi^2 +  (\nabla_b \Psi)(\nabla^b\Psi) - m_\Psi^2 \Psi^2\right)
  		\end{equation}
  		for the uncoupled combination. Here, $\Phi$ will be the system field and $\Psi$ the probe field, with respective masses $m_\Phi$ and $m_\Psi$. A simple interaction term that couples them together in a local region is  
  		\begin{equation}
  		S_{\text{int}}= - \int_\Mb\dvol\, \rho \Phi  \Psi,
  		\end{equation}
  		where $\rho$ is a real, smooth function compactly supported in compact region $K$.  
  		It is convenient to write the field equations in a matrix form
  		\begin{equation}\label{eq:eom}
  		T \begin{pmatrix}
  		\Phi\\ \Psi
  		\end{pmatrix} := \begin{pmatrix}
  		P & R \\ R & Q
  		\end{pmatrix}\begin{pmatrix}
  		\Phi\\ \Psi
  		\end{pmatrix}
  		=0,
  		\end{equation}
  		where $P=\Box+ m_\Phi^2$, $Q=\Box+m_\Psi^2$ are the free Klein--Gordon operators for the two fields and $R\Phi=\rho\Phi$. Writing the advanced $(-)$ and retarded $(+)$ Green operators for $P$, $Q$ and $T$ by $E_P^\pm$ and so forth, the Green functions of $T$ may be established via a Born series
  		\begin{equation}
  		E^\pm_T = 
  		\sum_{r=0}^\infty (-1)^r \begin{pmatrix}
  		E^\pm_P & 0 \\ 0 & E^\pm_Q 
  		\end{pmatrix} 
  		\left[\begin{pmatrix}
  		0 & R \\ R & 0
  		\end{pmatrix}\begin{pmatrix}
  		E^\pm_P & 0 \\ 0 & E^\pm_Q 
  		\end{pmatrix}\right]^r,
  		\end{equation} 
  		which converges at least for sufficiently weak coupling $\rho$~\cite{FewVer:20xx}. Thus, the (un)coupled classical field equation is Green hyperbolic\index{Green hyperbolic operator}~\cite{Baer:2015}. 
  		
  		The (un)coupled theory may be quantized by standard means; evidently the uncoupled theory has as its algebra the algebraic tensor product of the algebras for the free scalar fields of masses $m_\Phi$ and $m_\Psi$ respectively. The generators of these algebras will be denoted $\Phi(f)$ and $\Psi(f)$, labelled by test functions on $\Mb$.
  		It is of particular relevance to find the scattering morphism acting on elements
  		$\II\otimes\Psi(h)$ in the uncoupled algebra. In the case where $h$ is supported in the out region $M^+$, there is a simple formula 
  		\begin{equation}\label{eq:Theta_act}
  		\Theta(\II\otimes \Psi(h)) = \Phi(f^-)\otimes\II + \II\otimes \Psi(h^-),
  		\end{equation}
  		where
  		\begin{equation}\label{eq:fin}
  		\begin{pmatrix}  f^- \\ h^- \end{pmatrix}
  		=\begin{pmatrix} 0 \\ h \end{pmatrix}
  		-\begin{pmatrix} 0 & R \\ R & 0 \end{pmatrix}   E_T^-
  		\begin{pmatrix} 0 \\ h \end{pmatrix} \qquad (h\in\CoinX{M^+}).
  		\end{equation}
  		As $\Theta$ is a homomorphism, we immediately obtain the identity
  		\begin{equation}
  		\Theta (\II\otimes e^{i\Psi(h)}) = e^{i\Phi(f^-)}\otimes e^{i\Psi(h^-)}
  		\end{equation}
  		between formal power series in $h\in\CoinX{M^+}$, from which the induced system observables corresponding to probe observables $\Psi(h)^n$ may be computed. Indeed, the definition~\eqref{eq:inducedobs} allows
  		us to determine the generating function
  		\begin{equation}\label{eq:genfn_equiv}
  		\varepsilon_\sigma(e^{i\Psi(h)}) = \eta_\sigma(\Theta(\II\otimes e^{i\Psi(h)}))
  		=
  		\sigma(e^{i\Psi(h^-)}) e^{i\Phi(f^-)}
  		\end{equation}
  		for any probe preparation state $\sigma$, with $f^-$ and $h^-$ as before. 
  		Specific induced observables may be obtained by differentiation. For example, 
  		\begin{align}\label{eq:linear_equiv}
  		\varepsilon_\sigma(\Psi(h)) &= \Phi(f^-) + \sigma(\Psi(h^-))\II \\
  		\label{eq:quadratic_equiv}
  		\varepsilon_\sigma(\Psi(h)^2) &= \Phi(f^-)^2 + \sigma(\Psi(h^-)^2)\II
  		\end{align}
  		and so on.  A point of interest here is that none of the 
  		computations requires a Hilbert space representation -- everything takes place
  		at the level of the algebras. 
  		
  		A number of general features become apparent in this model. We give two examples. 
  		First, recall that the actual experiment takes place on the coupled system and that probe observable $\Psi(h)$ is identified at late times with observable $\widetilde{\Psi(h)}$ of the coupled system, while the system state is identified with a state $\utilde{\omega}_\sigma$. A simple calculation with the generating functions gives
  		\begin{equation}\label{eq:convolution}
  		 \utilde{\omega}_\sigma(e^{i\widetilde{\Psi(h)}})  = \sigma(e^{i\Psi(h^-)}) \omega(e^{i\Phi(f^-)}).
  		 \end{equation}
  		Given sufficient regularity, $\lambda\mapsto \omega(e^{i\lambda\Phi(f^-)})$ and 
  		 $\lambda\mapsto \sigma(e^{i\lambda\Psi(h^-)})$ are the characteristic 
  		 functions of probability measures for measurement outcomes of $\Phi(f^-)$ in state $\omega$ and $\Psi(h^-)$ in state $\sigma$. Using~\eqref{eq:convolution} we see that the  measurement outcomes of $\widetilde{\Psi(h)}$ in state $\utilde{\omega}_\sigma$ -- the outcomes of the actual experiment performed -- are distributed according to the convolution of these measures, with 
  		 characteristic function 
  		 $\lambda\mapsto \utilde{\omega}_\sigma(e^{i\lambda\widetilde{\Psi(h)}})$. This illustrates the general fact that the actual measurement is less sharp than the hypothetical measurement of the induced observable, due to fluctuations in the probe system. In particular, one has 
  		\begin{equation}
  		\Var(\widetilde{\Psi(h)};\utilde{\omega}_\sigma) 
  		= \Var(\Phi(f^-);\omega)+\sigma(\Psi(h^-)^2)
  		\end{equation}
  		for the variance of the measured observable, assuming for simplicity that $\sigma$ has vanishing one-point function. 
  		
  		Second, the localisation of the induced observables can be determined. Taking
  		the probe observable $\Psi(h)$ as our example (with $h\in\CoinX{M^+}$), the induced observable has localisation determined by the support of $f^-$, contained within the intersection $J^-(\supp h)\cap \supp\rho$. Unsurprisingly, the localisation of the probe observable must intersect the causal future of the coupling in order to constitute a nontrivial system measurement.  
  		
  		The general formalism allows one to assert that the induced observable may be localised roughly within the causal hull of $J^-(\supp h)\cap \supp\rho$. It is very tempting to try to ascribe it the localisation $\supp f^-$, particularly if $\rho$ were localised in a thin tube near a timelike curve, for the spatial dimensions of the causal hull are proportional (with constant equal to the speed of light) to the temporal duration of the coupling and it might be convenient to replace this with a much smaller spacetime volume. A simple argument can be given to disabuse the reader of such temptations.  Consider two induced observables, $\varepsilon_\sigma(\Psi(h))$ and $\varepsilon_\sigma(\Psi(h'))$. Their commutator may be computed as 
  		\begin{align}\label{eq:comm}
  		[\varepsilon_\sigma(\Psi(h)),\varepsilon_\sigma(\Psi(h'))] &= 
  		[\Phi( f^-),\Phi( f^{\prime -})] 
  		= 
  		iE_P( f^-,  f^{\prime -})\II ,
  		\end{align}
  		where
  		\begin{equation}
  		E_P(f_1,f_2) = \int_\Mb \dvol_\Mb f_1 \left(E^-_P f_2 - E^+_P f_2\right).
  		\end{equation}
  		Crucially, the right-hand side of~\eqref{eq:comm} depends on the geometry throughout the region 
  		\begin{equation}
  		S = (J^+(\supp f^-)\cap J^-(\supp f^{\prime -} ))\cup (J^-(\supp f^-)\cap J^+(\supp f^{\prime -} )).
  		\end{equation}
  		Typically, it will be possible to find $h'$ so that the supports of $f^{\prime -}$ and $f^-$ are equal, whereupon $S$ is the causal hull of $\supp f^-$. Consequently, there are questions concerning the induced observable $\varepsilon_\sigma(\Psi(h))$, e.g., it compatibility or otherwise with another observable, that 
  		cannot be answered without knowledge of the geometry of the causal hull of $\supp f^-$. 
  		It seems unhelpful or misleading, therefore, to ascribe any smaller localisation to $\varepsilon_\sigma(\Psi(h))$.

  		For example, if the coupling is supported precisely on a timelike curve segment $\gamma:[0,\tau]\to M$, then the localisation region must include $J^+(\gamma(0))\cap J^-(\gamma(\tau))$. For a uniformly accelerated trajectory of infinite proper duration, 
  		as in the Unruh--deWitt detector analysis, the localisation region becomes an entire wedge region of infinite spatial extent.
  		
  		\section{Conclusion}
  		
  		The framework set out here, and in full detail in~\cite{FV:2018}, provides a fully covariant measurement scheme for general QFTs in curved spacetimes, which brings together quantum measurement theory and algebraic QFT. It is not tied to particular models, and formulates its assumptions in abstract terms; on the other hand, it allows for practical computations in specific models. It describes both the correspondence between observables of the probe and induced system observables, and also the updating of states by post-selection, with non-selective measurement as a special case. Several more topics are discussed in~\cite{FV:2018} but not here, including the role of gauge invariance and symmetries, and a perturbative treatment of the specific model studied in~\ref{sec:model}. It was also shown -- for our model -- how the product on the probe algebra could be deformed to make the mapping $\varepsilon_\sigma$ into a homomorphism, providing a sense in which the system observables are partially represented within the probe algebra. 
  		
  		Above all, we wish to emphasise three points in particular. First, the localisation properties of the induced observables were discussed; they may all be localised within any connected open causally convex neighbourhood of the coupling region $K$. Such neighbourhoods necessarily contain the causal hull of $K$ and we have argued in the context of our model that no smaller localisation region can be expected. Second, the post-selected states satisfy an important consistency condition that allows multiple measurements to be combined into overarching measurements whenever they are subject to a causal order (and independently of the choice of order where relevant). Finally, incorporating the central insight of AQFT, our approach puts the principle of locality at its centre.  

  		\begin{acknowledgement} 
  			I thank Rainer Verch for useful comments on the text.
  		\end{acknowledgement}

\ifthenelse{\boolean{arXiv}}{%
}{%
\bibliographystyle{spmpsci}
\bibliography{covariant} 
\printindex
}
\end{document}